\documentclass[aps,showpacs,
amssymb,endfloats]{revtex4}
\usepackage{bm,amsfonts}
\begin{document}

\title {Light storing in a medium of atoms in the tripod configuration}
\author{A. Raczy\'nski}\email{raczyn@phys.uni.torun.pl}\author{M. Rzepecka}\author{J. Zaremba}

\affiliation{Instytut Fizyki, Uniwersytet Miko\l aja Kopernika,
ulica Grudzi\c{a}dzka 5, 87-100 Toru\'n, Poland,}
\email{raczyn@phys.uni.torun.pl}

\author{S. Zieli\'nska-Kaniasty}
\affiliation{Instytut Matematyki i Fizyki, Akademia
Techniczno-Rolnicza, Aleja Prof. S. Kaliskiego 7, 85-796
Bydgoszcz, Poland.}

\begin{abstract}

Light propagation and storing in a medium of atoms in the tripod
configuration driven by two control pulses are investigated
theoretically in terms of two polaritons and numerically. It is
shown that a magnetic field switched on at the pulse storage stage
changes the phase relations between the atomic coherences due to
the stored pulse, which leads to an essential modification of the
released pulse. Quantitative relations concerning the released
pulse and the coherences are given. A general situation when the
two control fields are not proportional at the pulse release stage
is also examined. It is shown that in both cases a single dark
state polariton is not sufficient to account for the pulse
evolution, which is connected with the fact that a part of the
signal remains in the medium after the release stage.
\\ 
\end{abstract}
\maketitle
\newpage
\section{Introduction}
A propagation of a weak laser pulse in atomic media driven by
laser control pulses has recently been a subject of intensive
investigations due to both the problem's fundamental character in
quantum optics as well as possible future applications. In
particular a significant pulse slowdown or its storing in the
medium have been achieved experimentally and interpreted
theoretically \cite{a1,a2,a3,a4,a5,a51}. This was first done by
controlling in time the electromagnetically induced transparency
(EIT) \cite{a6} in a medium of atoms in the so-called lambda
configuration including three active levels: the initial ground
state is coupled by a signal pulse with an upper state which is
coupled by the control field with another low-lying state.
Switching the control field off after the signal has entered the
medium results in light storing in the form of an atomic coherence
between the two lower states. Switching the control field on after
some time leads to a release of the trapped pulse. Light
propagation, storing and release can be described in terms of a
dark-state polariton, being a joint excitation (quasiparticle) of
the atom+field system \cite{a8} .

An interesting question, also from the practical point of view,
concerns a possibility of modifying the pulse at the storing
stage, which would enable one to process the information conveyed
by the light. This would require influencing the atomic coherence
in which the pulse has been written. The first experimental
attempt was to additionally switch on a magnetic field in the
direction of the pulse propagation \cite{a7}. This led to a
controlled change of the phase of the coherence corresponding to
the stored pulse and as a consequence to the respective
modification of the overall phase of the released pulse: the
latter was made to interfere with a reference pulse (which passed
by the medium) and the interference pattern was indeed observed.

Extending the atom+field system to include four active states
gives new possibilities to control light propagation and storing.
Coupling any of the lower states of the lambda system with a
fourth state allows one to temporarily prevent the signal from
being released or to release a signal of a different frequency
than that of the stored signal \cite{a9}. On the other hand, in
the double lambda configuration (with two upper states) it is
possible, after fulfilling a certain condition, to make the medium
transparent simultaneously for two pulses and, after storing them
in the form of a single atomic coherence between the two lower
states, one can turn the coherence in a controlled way into two
pulses of different frequencies \cite{a10,a11,a12,a13}.

Another atomic four-level system, which however has not been
exploited in the context of light storing, is that in a tripod
configuration, i.e. one including three low-lying states coupled
with a single upper state \cite{a14,a15}. Light propagation in the
generalized case of EIT and its slowdown in such an atomic medium
have been studied in Refs \cite{a16,a17}; however those works do
not concern light storing. Note that, contrary to the case of
lambda systems, for the tripod configuration a single pulse is
stored in the form of two atomic coherences. This enriches the
dynamics and open new possibilities of a coherent control of light
pulses. In particular light releasing in the case of a tripod
system may be performed at a few stages and in a more flexible way
than for a double lambda system. One also obtains new
possibilities of processing the stored light by changing the
relations between the two atomic coherences at the storing stage
using an additional interaction; this is more elaborate than
manipulating only a single coherence as in Ref. \cite{a7}.

In the present paper we theoretically investigate light
propagation and storing in a medium of atoms in the tripod
configuration. In the next section we first present the system of
the Maxwell-Bloch equations and then we discuss their approximate
solutions in terms of two polaritons. An evolution due to an
additional interaction at the storage stage, which does not cause
transitions but changes the phases of the coherences, cannot be
described in terms of the polaritons but it creates new initial
conditions for the polaritons' evolution at the release stage. For
proportional control pulses we give simple analytical predictions
of the height of the released pulse and for the values of the
coherences which will remain trapped in the medium. Section III
includes a numerical illustration of the theoretical results. In
particular we verify the predictions of the previous section in
the case in which the additional interaction is due to a constant
magnetic field parallel to the direction of propagation. We also
demonstrate the pulse evolution if at the release stage the two
control pulses are not switched on simultaneously.

\section{Theory}
Consider a four-level atomic model in the tripod configuration
(see Fig. \ref{fig1}) with a single upper state $a$ and three
lower states $b$ (initially populated), $c$ and $d$. The lower
states are coupled with the upper state by three co-propagating
fields: a weak signal field 1 of the envelope $\epsilon_{1}$ and
the frequency $\omega_{1}=E_{a}-E_{b}+\Delta_{1}$ and two stronger
control fields 2 and 3 of the envelopes $\epsilon_{2}$ and
$\epsilon_{3}$ and frequencies $\omega_{2}=E_{a}-E_{c}+\Delta_{2}$
and $\omega_{3}=E_{a}-E_{d}+\Delta_{3}$. Similarly as in earlier
papers dealing with similar subjects, propagation effects for the
control fields will be neglected. The density matrix $\rho$
fulfills the von Neumann equation completed with phenomenological
relaxation terms describing relaxation within the system. If we
transform-off the terms rapidly oscillating in time and make the
rotating wave approximation we obtain the following equations for
the density matrix $\sigma$
($\sigma_{ab}=\rho_{ab}\exp[i\phi_{1}]$,
$\sigma_{ac}=\rho_{ac}\exp[i\phi_{2}]$,
$\sigma_{ad}=\rho_{ad}\exp[i\phi_{3}]$,
$\sigma_{bc}=\rho_{bc}\exp[-i(\phi_{1}-\phi_{2})]$,
$\sigma_{bd}=\rho_{bd}\exp[-i(\phi_{1}-\phi_{3})]$,
$\sigma_{cd}=\rho_{cd}\exp[-i(\phi_{2}-\phi_{3})]$,
$\sigma_{ii}=\rho_{ii}$, where $\phi_{j}\equiv\omega_{j}t-k_{j}z$)

\begin{eqnarray}
i\dot{\sigma}_{aa}&=&
-\Omega_{1}^{*}\sigma_{ba}+\Omega_{1}\sigma_{ab}
-\Omega_{2}^{*}\sigma_{ca}+\Omega_{2}\sigma_{ac}
-\Omega_{3}^{*}\sigma_{da}+\Omega_{3}\sigma_{ad}
-i\gamma\sigma_{aa},\nonumber\\
i\dot{\sigma}_{bb}&=&\Omega_{1}^{*}\sigma_{ba}-\Omega_{1}\sigma_{ab}
+i\Gamma_{ab}\sigma_{aa},\nonumber\\
i\dot{\sigma}_{cc}&=&\Omega_{2}^{*}\sigma_{ca}-\Omega_{2}\sigma_{ac}
+i\Gamma_{ac}\sigma_{aa},\nonumber\\
i\dot{\sigma}_{dd}&=&\Omega_{3}^{*}\sigma_{da}-\Omega_{3}\sigma_{ad}
+i\Gamma_{ad}\sigma_{aa},\nonumber\\
i\dot{\sigma}_{ab}&=&(\Delta_{1}-\frac{i\gamma}{2})\sigma_{ab}
-\Omega_{1}^{*}(\sigma_{bb}-\sigma_{aa})-\Omega_{2}^{*}\sigma_{cb}
-\Omega_{3}^{*}\sigma_{db},\\
i\dot{\sigma}_{ac}&=&(\Delta_{2}-\frac{i\gamma}{2})\sigma_{ac}
-\Omega_{2}^{*}(\sigma_{cc}-\sigma_{aa})-\Omega_{1}^{*}\sigma_{bc}
-\Omega_{3}^{*}\sigma_{dc},\nonumber\\
i\dot{\sigma}_{ad}&=&(\Delta_{3}-\frac{i\gamma}{2})\sigma_{ad}
-\Omega_{3}^{*}(\sigma_{dd}-\sigma_{aa})-\Omega_{1}^{*}\sigma_{bd}
-\Omega_{2}^{*}\sigma_{cd},\nonumber\\
i\dot{\sigma}_{bc}&=&(\Delta_{2}-\Delta_{1})\sigma_{bc}-\Omega_{1}
\sigma_{ac}+\Omega_{2}^{*}\sigma_{ba},\nonumber\\
i\dot{\sigma}_{bd}&=&(\Delta_{3}-\Delta_{1})\sigma_{bd}-\Omega_{1}
\sigma_{ad}+\Omega_{3}^{*}\sigma_{ba},\nonumber\\
i\dot{\sigma}_{cd}&=&(\Delta_{3}-\Delta_{2})\sigma_{cd}-\Omega_{2}
\sigma_{ad}+\Omega_{3}^{*}\sigma_{ca}.\nonumber
\end{eqnarray}
In the above equations $\Gamma_{aj}$ is the transition rate from
$a$ to $j$ due to the spontaneous emission,
$\gamma=\Gamma_{ab}+\Gamma_{ac} +\Gamma_{ad}$, relaxation is due
solely to the latter process,
$\Omega_{1}(z,t)=\frac{1}{2\hbar}d_{ba}\epsilon_{1}(z,t)$,
$\Omega_{2}(t)=\frac{1}{2\hbar}d_{ca}\epsilon_{2}(t)$,
$\Omega_{3}(t)=\frac{1}{2\hbar}d_{da}\epsilon_{3}(t)$ are the Rabi
frequencies corresponding to the particular couplings, $d_{ij}$
being the dipole transition moments.

The above Bloch equations are accompanied by the Maxwell propagation equation for
the field 1, which reads in the slowly varying envelope approximation
\begin{equation}
(\frac{\partial}{\partial t}+c\frac{\partial}{\partial z})\Omega_{1}=
-i\kappa^{2}\sigma_{ba},
\end{equation}
where
$\kappa^{2}=\frac{N|d_{ab}|^{2}\omega_{1}}{2\epsilon_{0}\hbar}$,
$N$ being the atom density and $\epsilon_{0}$ - the vacuum
electric permittivity.

As in earlier papers, a simplified discussion can be performed in
the perturbative approximation with respect to the signal field 1
(but not with respect to control fields), in the resonance
conditions, with relaxation neglected, in the adiabatic
approximation, i.e. with $\dot{\sigma}_{ab}=0$. The Maxwell-Bloch
equations are then reduced to the form
\begin{eqnarray}
\Omega_{1}=-\Omega_{2}\sigma_{bc}-\Omega_{3}\sigma_{bd},\nonumber\\
\frac{1}{\Omega_{2}^{*}}\dot{\sigma}_{bc}=
\frac{1}{\Omega_{3}^{*}} \dot{\sigma}_{bd},\\
(\frac{\partial}{\partial t}+c\frac{\partial}{\partial
z})\Omega_{1}=\kappa^{2}
\frac{1}{\Omega_{2}^{*}}\dot{\sigma}_{bc}\nonumber.
\end{eqnarray}
It is convenient to perform an analysis of the solutions in terms of the
dark state polariton
\begin{equation}
\Psi=\exp(-i\chi)[\cos\theta \Omega_{1}-\kappa \sin\theta
(\cos\phi\exp(i\chi_{2}) \sigma_{bc}+\sin\phi\exp(i\chi_{3})
\sigma_{bd})],
\end{equation}
and another polariton
\begin{equation}
Z=[\sin\phi\exp(-i\chi_{3})\sigma_{bc}-\cos\phi\exp(-i\chi_{2})
\sigma_{bd}].
\end{equation}
We have set $\Omega^{2}\equiv |\Omega_{2}|^{2}+|\Omega_{3}|^{2}$,
$\tan\phi= |\Omega_{3}|/|\Omega_{2}|$,
$\chi_{2,3}=\arg(\Omega_{2,3})$, $\tan\theta\equiv\kappa/\Omega$.
The phase $\chi$ is defined by the relation
\begin{equation}
\dot{\chi}=\sin^{2}\theta(\cos^{2}\phi
\dot{\chi}_{2}+sin^2\phi\dot{\chi}_{3}).
\end{equation}
Eq. (4) is a natural generalization of the form of dark state
polaritons used for single and double lambda systems, in our case
of a single signal and two atomic coherences. The phase factors
have been singled out to admit complex control fields (note that
this variation of the problem has not been discussed before even
for a simple lambda system). The polariton $Z$ (Eq. (5)) is the
superposition of the atomic coherences orthogonal to that of Eq.
(4).

The two new variables satisfy the equations
\begin{eqnarray}
\frac{\partial}{\partial
t}\Psi+c\cos^{2}\theta\frac{\partial}{\partial z}\Psi= \nonumber\\
\exp[i(\chi_{2}+\chi_{3}-\chi)] \tan^{2}\theta\cos\theta
[\dot{\phi}+i\sin\phi\cos\phi(\dot{\chi}_{3}-\dot{\chi}_{2})]
\Omega Z,
\\ \frac{\partial}{\partial t} Z=
[-i(\cos^{2}\phi\dot{\chi}_{2}+
\sin^{2}\phi\dot{\chi}_{3})]Z+\nonumber\\
\exp[i(\chi-\chi_{2}-\chi_{3})]\cos\theta[-\dot{\phi}+i\sin\phi\cos\phi
(\dot{\chi}_{3}-\dot{\chi}_{2})]\frac{1}{\Omega}\Psi\nonumber.
\end{eqnarray}
In the situation when the phases $\chi_{2,3}$ of the control
fields are constant and equal to zero the equations simplify to
the form
\begin{eqnarray}
\frac{\partial}{\partial
t}\Psi+c\cos^{2}\theta\frac{\partial}{\partial z}\Psi=
\tan^{2}\theta\cos\theta \dot{\phi}\Omega Z,
\\ \frac{\partial}{\partial t} Z=
-\cos\theta \dot{\phi}\frac{1}{\Omega}\Psi\nonumber.
\end{eqnarray}

When $\phi=const.$, which means that the control fields change at
the same rate, the two equations are decoupled. If additionally
$Z=0$ initially, the evolution can be described by a single dark
state polariton, similarly as in the case of a $\Lambda$ system
with a properly chosen control field $\Omega$:
$\Psi(z,t)=\Psi(z-\int_{0}^{t}c\cos^{2}\theta(t)dt,0)$. On the
other hand if $\dot{\phi}\neq 0$ or if the coherences at the
storing stage have been somehow changed so that $Z\neq0$, the
further evolution must be described in terms of both polaritons
and in general not all the excitation will leave the medium.

Let us now admit the simplest possible manipulation performed on
the pulse at the storage stage: we modify the phases of the
coherences $\sigma_{bc}$ and $\sigma_{bd}$ by switching on an
interaction $U$which is diagonal in the basis $(a, b ,c, d)$. The
levels will be then shifted and the two coherences will acquire
additional phase shifts. Assume that a pulse has been stopped by a
simultaneous switch-off of the two control fields and the
coherences at the time $t_{0}$ have the values
$\sigma_{bc}^{0}(z)$ and $\sigma_{bd}^{0}(z)$. Then the
interaction $U$ is switched on. At this stage the evolution cannot
be simply described in terms of polaritons, because the resonance
is spoilt due to the level shifts. Assume for simplicity that the
additional interaction has modified only the phase of
$\sigma_{bc}$. At the time $t_{1}$ after the interaction $U$ has
been switched off but before switching the control fields 2 and 3
on, the coherences are
$\sigma_{bc}(z,t_{1})=\sigma_{bc}^{0}(z)\exp(i\delta)$ and
$\sigma_{bd}(z,t_{1})=\sigma_{bd}^{0}(z)$, $\delta$ being the
pulse area corresponding to $U$. Those values of the coherences
are in general not suitable for the evolution to be described by a
dark state polariton only; in particular it is not so even if the
control fields which will release the pulse are the replicas of
those used to store it. However, each of the two coherences can be
decomposed into two parts, one of which (marked by ') will
contribute to the dark state polariton and the other (marked by
'') - to the excitation $Z$. The following equations will be
satisfied during the pulse release stage ($t>t_{1}$), again with
$\dot{\phi}=0$,
\begin{eqnarray}
\sigma_{bd}^{0}(z)=\tan\phi\sigma_{bc}^{0}(z),\\
\sigma_{bd}(z,t)-\sigma_{bd}^{0}(z)=\tan\phi
[\sigma_{bc}(z,t)-\sigma_{bc}^{0}(z)\exp(i\delta)],
\\ \sigma_{bc}(z,t)=\sigma_{bc}'(z,t)+\sigma_{bc}''(z,t),\\
\sigma_{bd}(z,t)=\sigma_{bd}'(z,t)+\sigma_{bd}''(z,t),\\
\cos\phi\sigma_{bc}''(z,t)+\sin\phi\sigma_{bd}''(z,t)=0,\\
\sin\phi\sigma_{bc}'(z,t)-\cos\phi\sigma_{bd}'(z,t)=0.
\end{eqnarray}
The solutions $\sigma''$ of the above equations are
\begin{eqnarray}
\sigma_{bc}''(z,t)=\sin^{2}\phi
[\exp(i\delta)-1]\sigma_{bc}^{0}(z)=const.,\\
\sigma_{bd}''(z,t)=-\cos^{2}\phi
[\exp(i\delta)-1]\sigma_{bd}^{0}(z)=const.
\end{eqnarray}
At the time $t_{1}$ the solutions $\sigma'$, the dark state
polariton and the Z excitation are
\begin{eqnarray}
\sigma_{bc}'(z,t_{1})=\sigma_{bc}^{0}(z)[\cos^{2}\phi\exp(i\delta)+\sin^{2}\phi],\\
\sigma_{bd}'(z,t_{1})=\sigma_{bd}^{0}(z)[\cos^{2}\phi\exp(i\delta)+\sin^{2}\phi].
\end{eqnarray}
\begin{eqnarray}
\Psi(z,t_{1})=-\kappa(\cos\phi\sigma_{bc}'(z,t_{1})+\sin\phi\sigma_{bd}'(z,t_{1})),\\
Z(z,t_{1})=\sin\phi\sigma_{bc}''(z,t_{1})-\cos\phi\sigma_{bd}''(z,t_{1}).
\end{eqnarray}
Note that for $t>t_{1}$ $Z(t)=Z(t_{1})$. This means that the part
of the atomic excitation constituting the polariton $Z$ does not
leave the medium and preserves its shape, while the other part is
turned back into the pulse and leaves the medium. The height and
phase of the leaving pulse can be obtained from the equation
\begin{equation}
\Omega_{1}(z,t)=\Psi(z,t)\cos\theta(t),
\end{equation}
where $\Psi$ has the shape given by Eq. (19) and it has moved from
the initial position by $\int_{0}^{t}c \cos^{2}\theta(t)dt$.

\section{Numerical examples}
In this section we demonstrate numerical results of the pulse
storing and release in two cases in which the polariton $Z$ is
engaged: (1) the case of both control fields being switched on and
off simultaneously at the same rate with the interaction $U$
present at the storage stage and (2) the case of a time delay
between the instants of the switch-on of the control fields. We
have numerically solved the Maxwell-Bloch equations (Eqs (1) and
Eq. (2)) for the tripod system in the moving reference frame
($\xi=z$, $\tau=t-z/c$). Resonance has been assumed for all three
transitions. The upper state $a$ has a width of $3\Gamma$ due to a
spontaneous emission of rate $\Gamma= 4\times 10^{-10}$ a.u. (2.6
MHz) to all three lower states. The relaxation rates for the
coherences have been taken half of that of the upper state (as in
the spontaneous emission). The entering signal pulse $\Omega_{1}$,
shaped as a sine square, had the maximum value of 0.025$\Gamma$
while the control fields, taken real and equal, both switched on
and off as a hyperbolic tangent, had maximum values of 5 $\Gamma$,
which was enough to make our probe transparent. The length of the
probe was of order of 1 cm. The initial width of the signal pulse
was 2.4 $\mu$s, while the time of light storing was of order of 10
$\mu$s. The atom density was $N=2\times 10^{12}$ cm$^{-3}$.

The first case concerned changing the phases of the coherences at
the storing stage. Its realization, for the levels engaged in the
transitions being Zeeman sublevels of given quantum numbers F and
M, may be carried out by switching on a magnetic field $B$
parallel to the propagation direction. Each of the levels is
shifted by $\Delta E= g_{F}BM/2$, $g_{f}$ being the Lande factor.
The phases of the states $a-d$ and thus the phases of the
coherences $\sigma_{bc}$ and $\sigma_{bd}$ evolve according to Eqs
(1), with $\Omega_{1,2,3}=0$ and with modified detunings. If, for
example, in the case of $^{87}$Rb, the numbers (F,M) are (2,-1)
for the state $b$, (2,0) for $a$, (2,1) for $c$ and (1,1) for $d$,
$\Delta E_{b}=-B/4$, $\Delta E_{c}=B/4$, $\Delta E_{d}=-B/4$. This
means that $\sigma_{bd}$ will remain unchanged while $\sigma_{bc}$
will obtain an extra phase $\delta=-B\tau/2$, where $\tau$ is the
duration of the magnetic pulse. We have assumed rectangular pulses
of the magnetic field of duration of 2.4 $\mu$s and of the value
of the magnetic induction up to 3$\times10^{-5}$ T. The two
control pulses were identical during the signal storing and
release

Fig. \ref{fig2} presents the shape of the signal pulse at the end
of the sample (1 cm) after it has been stored and released. Its
height is a function of the area $\delta$ of the magnetic pulse.
For $\delta$ being a multiple of $2\pi$ the magnetic field does
not change the coherence $\sigma_{bc}$, so the analysis may be
performed in terms of the single dark state polariton $\Psi$ and
the released pulse has the same height as the stored one. For
other values of $\delta$ the release stage may be again discussed
in terms of uncoupled polaritons but with new initial conditions.
The coherences are split into two parts. The absolute value of the
part of both $\sigma_{bc}$ and $\sigma_{bd}$, which will enter
$\Psi$, is  $\sqrt{1-\sin^{2}2\phi \sin^{2}\frac{\delta}{2}}$
times smaller than its original value (cf. Eq. (17) and Eq. (18))
and so will be the height of the released pulse compared with that
of the stored one. Indeed, one can see in Fig. \ref{fig2} that for
$\phi=\pi/4$ the heights vary as $\cos\frac{\delta}{2}$. Fig.
\ref{fig3} shows the spatial distribution of the corresponding $Z$
polariton at the end of the evolution, i.e. after the released
pulse has left the medium. Its height varies as $\sin \delta/2$,
in agreement with Eqs (15), (16) and (20). These numerical results
confirm our quantitative predictions of the values of the heights
and phases of both the released pulse and that part of the
excitation which will stay inside the medium unless it is
destroyed by relaxation processes absent from our model.

The other case examined in this paper, in which the polariton $Z$
is engaged, is that in which the two control pulses are not
proportional at the release stage. In the situation of Figs
\ref{fig4} and \ref{fig5} the control fields were proportional at
the storage stage but the field 3 preceded the field 2 by 3.6
$\mu$s at the release stage: due to $\dot{\phi}\neq 0$ this leads
to the production of nonzero $Z$ (cf. Eqs (8)). Fig. \ref{fig4}
shows the space-time evolution of the signal pulse. One can see
the entering pulse at the left hand side. After the storage stage
the signal is released in two portions. The first portion is
released from the single coherence $\sigma_{bd}$ by the control
pulse 3, similarly as in a simple lambda system. Then the other
control pulse 2 causes a release of the signal stored in the
coherence $\sigma_{bc}$. However, the latter process occurs at the
presence of both control fields, which causes a difference of the
two parts of the signal, as concerns they heights and initial
velocities. The first part is released with a zero initial
velocity while the second part has a nonzero velocity from the
very beginning (see Fig. \ref{fig4}); after the two control fields
attain their final values the velocities become equal: the
structures in the figure become parallel for large $\tau$. To
obtain a full symmetry one should switch the first control field
off before switching the second one on. Then one would obtain two
identical released pulses shifted in time, as in two independent
simple lambda systems. Fig. \ref{fig5} shows the time evolution of
the polariton $Z$. One can see that it is equal to zero at the
storing stage, when $\dot{\phi}=0$. Later it becomes almost equal
to $\sigma_{bc}$ when the first pulse is generated from the
coherence $\sigma_{bd}$. Finally due to a coupling of the two
coherences there occurs their alignment, so that they have the
same absolute value and opposite signs. This polariton does not
leave medium at all. The proportions of the heights of the signal
pulses at the particular stages of release can be regulated by
changing the heights of the control pulses. However, it seems that
there are no simple formulas which would allow for quantitative
predictions, as in the first case discussed in this section.
\section{Conclusions}
Light propagation and storing in a medium of atoms in the tripod
configuration has been examined numerically and discussed in terms
of polaritons. The signal was trapped in the form of two atomic
coherences. In the very special situation in which the two control
pulses were proportional the pulse storing and release were
equivalent to those for a single lambda system with a properly
chosen control field. Two particular situations were investigated
in detail in which the usual dark state polariton was not
sufficient to describe the process and another polariton had to be
invoked. Both in the case in which the medium with a pulse stored
inside was subject to an interaction with a  magnetic field as
well as in the case in which the two control pulses were not
proportional, a part of the excitation remained trapped also after
the release phase. In the former case we have given simple
relations which allow one to predict the height of the released
pulse and the shape of the trapped excitation. We thus have a
rather simple means of processing information carried by a pulse:
not only can it be stored in the medium or made inaccessible in a
reversible way, but also the atomic excitation in which it has
been written, can be split into parts, of which the signal can be
read independently when required.
\begin{acknowledgments}
This work is a part of a program of the National Laboratory of AMO
Physics in Toru\'n, Poland.
\end{acknowledgments}

\newpage

\begin{figure}
\caption {\label{fig1} Color online. Atomic levels in the tripod
configuration.}
\end{figure}

\begin{figure}
\caption{\label{fig2} Color online. The time dependence of the
released pulse at the end of the sample as a function of the
magnetic pulse area.}
\end{figure}

\begin{figure}
\caption{\label{fig3} Color online. The spatial distribution of
the $Z$ polariton after the released pulse has left the sample as
a function of the magnetic pulse area.}
\end{figure}

\begin{figure}
\caption{\label{fig4}  Color online. The space-time course of the
pulse trapping and release in the case of the releasing control
fields shifted in time. }
\end{figure}

\begin{figure}
\caption {\label{fig5} Color online. The evolution of the
polariton $Z$ corresponding to the pulse of Fig. \ref{fig3}.}
\end{figure}

\end{document}